# Superconductivity at 1 K in $Cd_2Re_2O_7$

M. Hanawa, Y. Muraoka, T. Tayama, T. Sakakibara, J. Yamaura, and Z. Hiroi

*Institute for Solid State Physics, University of Tokyo, Kashiwanoha, Kashiwa, Chiba 277-8581, Japan*

We report the first pyrochlore oxide superconductor $Cd_2Re_2O_7$. Resistivity, magnetic susceptibility, and specific heat measurements on single crystals evidence a bulk superconductivity at 1 K. Another phase transition found at 200 K suggests that a peculiar electronic structure lies behind the superconductivity.

Interplay between localized and itinerant electrons has been one of the most exciting subjects in solid state physics. In particular, various transition-metal (TM) oxides which exist in the vicinity of a metal-insulator (MI) transition have been studied extensively, and many intriguing phenomena such as high-temperature superconductivity in cuprates and charge/orbital ordering in manganites or others have been found. They would further attract many physicists in the future because of huge family of compounds already known and still hidden under the iceberg.

Superconducting TM oxides known before the discovery of cupric oxide superconductors are rather limited. Only a few Ti, Nb, and W oxides which crystallize in the NaCl, spinel, or perovskite structure have been reported for $3d$, $4d$, and $5d$ series, respectively. The highest superconducting transition temperature $T_c$ among those non-cuprates was attained in a spinel compound $LiTi_2O_4$ ($T_c$ = 13.7 K) [1]. Recently, $Sr_2RuO_4$ is studied well as representing an unusual $p$-wave superconductivity at 1 K [2].

There is another class of compounds forming a large family of TM oxides which crystallize in the pyrochlore structure with a chemical formula $A_2B_2O_7$ where the B represents TMs [3]. However, no superconductivity has been observed there so far in spite of many metallic compounds present in the family. Looking in the general trend of electronic properties for pyrochlore compounds, most $3d$ and $4d$ TM pyrochlores are insulators owing to large electron correlations as well as relatively small electron transfers along the bent B-O-B bonds (110° ~ 140°). Molybdenum pyrochlores exist near the metal-insulator boundary, where ferromagnetic metals appear as the ionic radius of the counter cations is increased [4]. On one hand, metallic pyrochlores can be found, when additional electrons are supplied from A cations like typically in Mn and Ru pyrochlores such as $Tl_2Mn_2O_7$ and $Bi_2Ru_2O_7$ [5, 6]. In contrast, $5d$ TM pyrochlores are mostly metallic because of relatively spreading $5d$ orbitals [3]. A rare exception reported previously is found in $Os^{5+}$ pyrochlores like $Cd_2Os_2O_7$ [7, 8] and $Ca_2Os_2O_7$ [9], where a MI transition occurs with temperature. An $Os^{5+}$ ion has a $5d^3$ electron configuration and thus the $t_{2g}$ orbital is half-filled, suggesting a possible Mott-Hubbard type MI transition. Note that most of $Os^{4+}$ pyrochlores are metals. These examples illustrate that the effect of electron correlations is still important even for the $5d$ electron systems in the pyrochlore structure. We have searched for novel phenomena in pyrochlore compounds near $Cd_2Os_2O_7$ and reached to $Cd_2Re_2O_7$ which becomes the first superconductor in the pyrochlore family.

Another important feature on the pyrochlore oxide is magnetic frustration on the three-dimensional tetrahedral network sharing vertices[10]. When $d$ electrons are localized on the B site ions with interacting antiferromagnetically, classical Néel order is suppressed by the frustration and a quantum spin liquid state can be stabilized instead [11]. Even for ferromagnetic Ising spin systems on the A sublattice like $Dy_2Ti_2O_7$, it is shown that the effect of frustration leads to a highly degenerate ground state called 'spin ice' [12]. More interestingly, an exotic ground state is to be realized when $d$ electrons keep an itinerant character near a MI transition. It is clearly illustrated with a mixed-valent compound $LiV_2O_4$ which has the spinel structure comprising a similar frustrated lattice and exhibits an unusual "heavy-Fermion" behavior [13, 14]. Superconductivity seen in another spinel compound $LiTi_2O_4$ may be interesting in this context. However, it has not been studied in detail because of difficulty in preparing single crystals.

There are a few studies on $Cd_2Re_2O_7$. Donohue *et al.* prepared a single crystal, determined the crystal structure, and reported the resistivity which was metallic above 4 K [15]. Blacklock and White measured the specific heat above 1.8 K using a polycrystalline sample and found the Sommerfeld coefficient $\gamma$ to be 13.3 mJ/K$^2$ mol Re [16]. Since the $Cd^{2+}$ and $Re^{5+}$ have $4d^{10}$ and $4f^{14}5d^2$ outer electron configurations, respectively, only the $Re^{5+}$ is expected to underlie the electronic and magnetic properties of $Cd_2Re_2O_7$. It is known that $Re^{5+}$ is not present as a stable state in any binary or ternary oxide system [17]. Subramanian *et al.* suggested that more stable $Re^{6+}$ and $Re^{4+}$ are the entities instead of $Re^{5+}$ in $Cd_2Re_2O_7$ [3]. This suggests an inherent charge fluctuation present in this compound. We prepared single crystals of $Cd_2Re_2O_7$ and measured resistivity, magnetic susceptibility, and specific heat down to ~ 0.4 K. Surprisingly observed at 1-2 K are a sharp drop in resistivity, a large diamagnetic signal in magnetization, and a well-defined $\lambda$-type anomaly in specific heat. These experimental facts give a strong evidence for the occurrence of superconductivity. Moreover, the resistivity and susceptibility show peculiar behaviors at high temperatures, implying interesting normal-state properties lying behind the superconductivity.

Single crystals of $Cd_2Re_2O_7$ were prepared by assuming the chemical reaction, $2CdO + 5/3ReO_3 + 1/3Re \rightarrow Cd_2Re_2O_7$. Stoichiometric amounts of these starting powders were mixed in an agate mortar and pressed into a pellet. The reaction was carried out in an evaluated quartz ampoule at 800-900ºC for 70 h. The product appeared as purple octahedral crystals of a few mm on an edge which adhered to the walls of the ampoule. The crystals have grown probably by a vapor transport mechanism, as discussed in a previous study [15], because both $CdO$ (Cd) and $ReO_3$ ($Re_2O_7$) are volatile. The chemical composition of the crystals examined by an electron-probe microanalyser (EPMA) was $Cd/Re = 1.00 \pm 0.02$. The oxygen content was not determined in this study. The previous study using crystals prepared in similar conditions suggested that the oxygen nonstoichiometry was negligible [15]. A powder X-ray diffraction (XRD) pattern was taken at room temperature and indexed on the basis of a face-centered cubic unit cell with $a = 1.0226(2)$ nm, which is slightly larger than the previously reported value of 1.0219 nm.

Resistivity and specific heat measurements were carried out on many crystals in a Quantum Design PPMS system equipped with a $^3$He refrigerator down to 0.4 K. The former was measured by the standard four probe method using crystals of typical dimensions 2 mm $\times$ 0.5 mm $\times$ 0.1 mm. The current density for the measurements was about $10^4$ A/m$^2$. Specific heat was measured by the heat-relaxation method using crystals of 10-20 mg in weight. To perform dc magnetization measurements at very low temperature below 2 K, we used a Faraday-force capacitive magnetometer. A field gradient of 500 Oe/cm was applied to a crystal in addition to homogeneous external field. The detail of this technique was reported previously [18]. Magnetic susceptibility above 1.7 K was measured in a Quantum Design MPMS system.

When the crystals were cooled below 2 K, we found a very sharp drop in resistivity probably due to superconductivity, as typically shown for two crystals in Fig. 1(a). The resistivity below the critical temperature $T_c$ was nearly zero within our experimental resolution of ~ 10 nV for the voltage detection. The onset temperature is 1.45 (2.15) K and a zero-resistivity is attained below 1.30 (1.85) K for crystal A (B). We have carried out measurements on more than 10 crystals and found similar sharp drops for all of them. However, the $T_c$ values were scattered between 1 K and 2 K. When the magnetic field was applied, the transition curves shifted to lower temperatures systematically.

The temperature dependence of the resistivity at high temperatures are quite unusual as shown in Fig. 1(b). It is almost temperature independent around room temperature, while, with cooling down, it suddenly starts to decrease at about 200 K. Another small anomaly is seen around 120 K, where the curvature is slightly changed. The decrease in the resistivity tends to saturate at low temperature, where the temperature dependence is approximately proportional to $T^3$. The resistivity at 300 K and just above the $T_c$ are 640 (360) $\mu\Omega$cm and 21 (11) $\mu\Omega$cm for crystal A (B), respectively. The ratio is about 30. These resistivity values varied slightly from crystal to crystal, but the ratio was always nearly the same.

Associated with the observation of the zero-resistive transition, a large diamagnetic signal due to the Meissner effect was observed below 1 K as shown in Fig. 2. The hysteresis loop measured at 0.45 K is characteristic of a type-II superconductor with $H_{c1}$ less than 0.002 T and $H_{c2}$ of about 0.17 T, respectively. The magnetization of the peak top at 0.002 T is -0.19 emu/g, which corresponds to $M/H$ = -0.0095 emu/g. This value is close to a value of -0.009 emu/g expected for perfect exclusion of vortices in $Cd_2Re_2O_7$. Therefore, the superconducting volume fraction in our crystal is large enough to conclude a bulk property. The inset to Fig. 2 shows the magnetization measured with increasing and decreasing temperature. The applied field may be close to $H_{c1}$, but not exactly known. The $T_c$ is determined as 0.98 K, which is considerably lower than decided in the resistivity measurements.

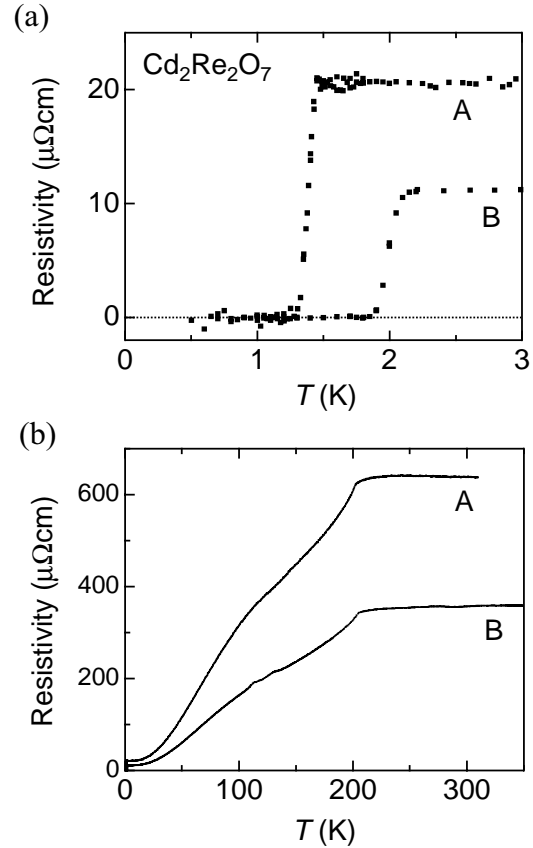

FIG. 1. Temperature dependence of resistivity for two $Cd_2Re_2O_7$ single crystals A and B. The measurements were carried out on cooling for crystal A and on heating for crystal B.

The superconducting transition was also detected by specific heat $C$ measurements. As shown in Fig. 3, a distinct $\lambda$-type anomaly is seen at zero field below 1 K, and it shifted to lower temperatures with increasing magnetic fields applied. The observed large jump at the transition evidences the bulk nature of superconductivity in the present compound. By fitting the data between 1 K and 10 K to the form $C = \gamma T + \alpha T^3 + \beta T^5$, we obtained $\gamma$ = 15.1 mJ/K$^2$ mol Re, $\alpha$ = 0.111 mJ/K$^4$ mol Re, and $\beta$ = $1.35 \times 10^{-6}$ mJ/K$^6$ mol Re. This $\gamma$ value is slightly larger than that reported previously [16]. It is rather small compared with the value of 37.5 mJ/K$^2$ mol reported for $Sr_2RuO_4$ [2]. The Debye temperature deduced from $\alpha$ is 458 K. The electronic specific heat $C_e$ was obtained by subtracting the lattice contribution estimated above, and $C_e/T$ is plotted in Fig.3. The $T_c$ determined from the midpoint of the jump in $C_e/T$ is 0.97 K, and the transition width between 25% and 75% height of the jump is 30 mK, which is much smaller than that reported for $Sr_2RuO_4$ [2]. The jump in $C_e$ at $T_c$ is 16.9 mJ/K mol Re, and thus $\Delta C_e/\gamma T_c$ is 1.15, which is considerably smaller than that expected for superconductivity with an isotropic BCS gap, 1.43. Specific heat data measured on other several crystals are essentially the same, giving nearly the same $T_c$, $\Delta T_c$, and jump at $T_c$. Since the data is limited above 0.4 K, it is difficult to discuss how the specific heat reaches zero as $T \rightarrow 0$. From the entropy balance, however, assuming that the normal-state specific heat is simply $\gamma T$, it is considered that the specific heat must decrease with temperature rather quickly as in an exponential form than in a power law as observed in $Sr_2RuO_4$ [2]. This suggests that the superconducting ground state of $Cd_2Re_2O_7$ is possibly nodeless unlike $Sr_2RuO_4$. Nevertheless, the gap can be anisotropic, because there is a significant difference between the observed and calculated specific heat data assuming an isotropic gap, as compared in Fig. 3. From the field dependence of $T_c$, the upper critical field $H_{c2}$ was roughly estimated to be 0.21 T, which corresponds to a coherence length of 40 nm.

The $T_c$ value is consistent between the magnetization and specific heat measurements, while the resistivity measurements gave much higher values of 1-2 K. This might be because the surface of crystals has been modified in some way to raise $T_c$. The former measurements probe the bulk superconductivity inside crystals, while the latter can be determined by the surface. We checked the surface of crystals by EPMA, but did observe neither deviation of the metal ratio nor segregation of other phases. One possible explanation would be a small change in oxygen content or metal ratio occurring at the crystal surface. Annealing experiments in various oxygen atmosphere are now in progress.

Magnetic susceptibility $\chi$ at high temperature is shown in Fig. 4, where a kink is seen at 200 K which coincides with the anomaly observed in the resistivity. The $\chi$ seems to show a broad, rounded maximum around room temperature and decreases rapidly below 200 K. The $\chi$ value at 5 K is reduced by 40 % from the maximum value at high temperature. There was little anisotropy in $\chi$ between two measurements where a magnetic field was applied along the <111> and <110> axes of the crystal.

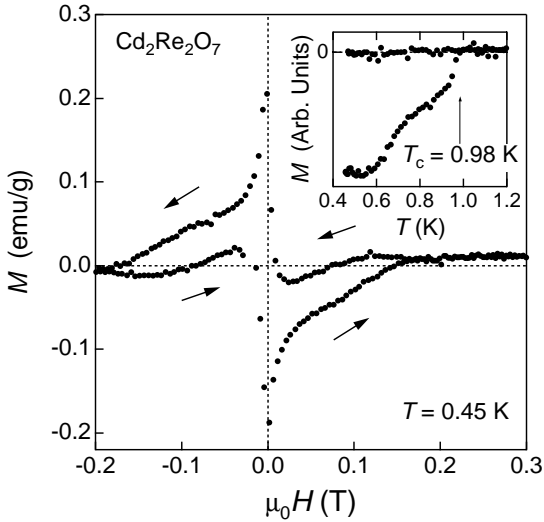

FIG. 2. Magnetization versus magnetic field curve measured at 0.45 K showing a hysteresis loop characteristic of a type-II superconductor. The temperature dependence shown in the inset exhibits $T_c$ = 0.98 K.

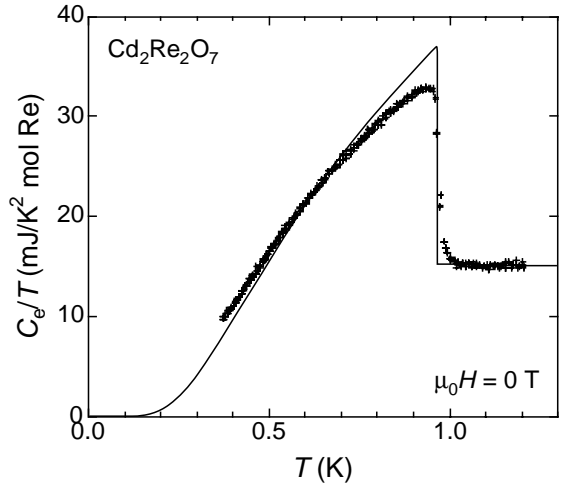

FIG. 3. Electronic specific heat $C_e$ divided by temperature for a $Cd_2Re_2O_7$ single crystal showing a $\lambda$-type anomaly associated with a superconducting transition at 0.97 K without magnetic fields. The solid line shows calculated $C_e/T$ assuming a superconducting transition with an isotropic BCS gap.

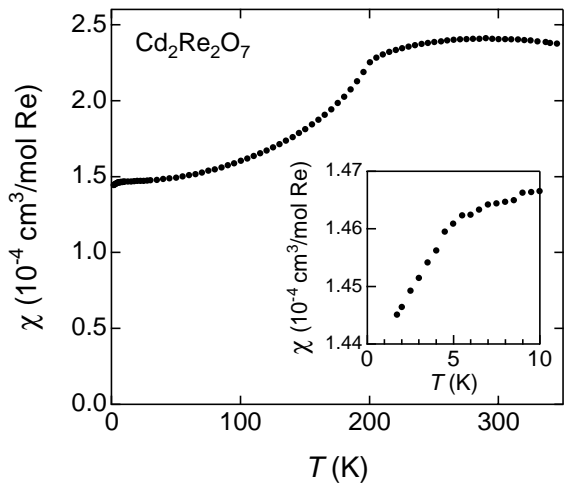

FIG. 4. Temperature dependence of magnetic susceptibility in a wide temperature range. The measurement was performed on heating in a magnetic field of 1 T applied nearly parallel to the <111> axis of the crystal. The inset shows an enlargement of the data at low temperature.

The Wilson ratio calculated from the $\chi$ value at 5 K and the $\gamma$ is 0.72, much smaller than unity expected for free electron gas. Below 5 K, there is a strange downturn in the $\chi$ curve as shown in the inset to Fig. 4. No thermal hysteresis was detected between heating and cooling for this crystal. The resistivity did not show any corresponding anomalies. The magnitude of this downturn was sample dependent.

Concerning the anomaly observed at 200 K, preliminary XRD, specific heat, and NMR experiments all suggested that there is a second-order phase transition without magnetic order. In particular, the XRD measurements indicated that there occurs a small structural change from cubic $Fd3m$ to another cubic $F\bar{4}3m$, implying a slight deformation for the oxygen octahedra as well as the tetrahedral network of Re ions. The detail will be reported elsewhere. Anyway, there is a distinct phase transition at 200 K in $Cd_2Re_2O_7$ which dramatically affects resistivity, magnetic susceptibility, and also crystal structure, and thus must induce a significant change in the electronic structure . It would be crucial for understanding the superconductivity in $Cd_2Re_2O_7$ to elucidate the essential feature of this phase transition at high temperature.

A $Re^{5+}$ ion in $Cd_2Re_2O_7$ has two $5d$ electrons which should occupy the $t_{2g}$ orbital. In the pyrochlore structure of space group $Fd3m$ a certain deformation of a $BO_6$ octahedron is allowed. In most compounds an octahedron is compressed along the <111> direction, while it is slightly elongated in $Cd_2Re_2O_7$. Then, it is considered that in $Cd_2Re_2O_7$ the threefold degenerate $t_{2g}$ level splits into $a_{1g}$ and $e_g$ levels, the latter having a lower energy. Therefore, a kind of multi-band character arising from these orbitals is expected near the Fermi level in the metallic state, as seen in $Sr_2RuO_4$. Preliminary band structure calculations by Harima suggested an interesting flat-band character just below the Fermi level [19]. The structural change at 200 K must influence this basic electronic structure.

In conclusion we found superconductivity in the pyrochlore compound $Cd_2Re_2O_7$. Although the nature of superconductivity is not clear at present, we expect a novel physics involved in this compound on the basis of electron correlations near the MI transition as well as frustration on the pyrochlore structure. In addition, the phase transition at high temperature is intriguing, which might be related to charge, spin, and orbital degrees of freedom. It is also important to investigate the other pyrochlore compounds which show metallic conductivity, in order to test whether a superconducting ground state is general in the pyrochlore oxides or $Cd_2Re_2O_7$ presents a special case.

We would like to thank T. Yamauchi, M. Takigawa, M. Imada, K. Ueda and Y. Ueda for helpful discussions during the course of this study, and Y. Ueda for the use of the SQUID magnetometer and F. Sakai for chemical analysis. This research was supported by a Grant-in-Aid for Scientific Research on Priority Areas (A) given by The Ministry of Education, Culture, Sports, Science and Technology, Japan.